\documentclass[final,3p,times,twocolumn]{elsarticle}
\usepackage{amssymb}
\usepackage{color}
\usepackage[nodots]{numcompress}
\usepackage[hyperindex,breaklinks]{hyperref}
\PassOptionsToPackage{linktocpage}{hyperref}
\usepackage{breakurl} 
\journal{Computers \& Fluids}

\begin{document}
\newcommand{\COMMENT}{\textcolor{red}}

\begin{frontmatter}

  \title{Performance engineering for the Lattice Boltzmann method on
    GPGPUs:\\ Architectural requirements and performance results}

\author[rrze]{J. Habich}
\ead{Johannes.Habich@rrze.uni-erlangen.de}
\author[inf]{C. Feichtinger}
\ead{Christian.Feichtinger@informatik.uni-erlangen.de}
\author[inf]{H. K\"ostler}
\ead{Harald.Koestler@informatik.uni-erlangen.de}
\author[rrze]{G. Hager}
\ead{Georg.Hager@rrze.uni-erlangen.de}
\author[rrze,inf]{G. Wellein}
\ead{Gerhard.Wellein@rrze.uni-erlangen.de}

\address[rrze]{Erlangen Regional Computing Center, University of Erlangen-Nuremberg, Germany}
\address[inf]{Department for Computer Science, University of Erlangen-Nuremberg, Germany}

\begin{abstract}
  GPUs offer several times the floating point performance and memory
  bandwidth of current standard two socket CPU servers, e.g. NVIDIA
  C2070 vs. Intel Xeon Westmere X5650. The lattice Boltzmann method
  has been established as a flow solver in recent years and was one of
  the first flow solvers to be successfully ported and that performs
  well on GPUs. We demonstrate advanced optimization strategies for a
  D3Q19 lattice Boltzmann based incompressible flow solver for GPGPUs
  and CPUs based on NVIDIA CUDA and OpenCL. Since the implemented
  algorithm is limited by memory bandwidth, we concentrate on
  improving memory access. Basic data layout issues for optimal data
  access are explained and discussed. Furthermore, the algorithmic
  steps are rearranged to improve scattered access of the GPU
  memory. The importance of occupancy is discussed as well as
  optimization strategies to improve overall concurrency. We arrive at
  a well-optimized GPU kernel, which is integrated into a larger
  framework that can handle single phase fluid flow simulations as
  well as particle-laden flows. Our 3D LBM GPU implementation reaches
  up to 650 MLUPS in single precision and 290 MLUPS in double
  precision on an NVIDIA Tesla C2070.
\end{abstract}


\end{frontmatter}

% \linenumbers

%%%% MAIN PART
%%%%%%%%%%%%%%%%%%%%%%%%%%%%%%%%%%%%%%%%%%%%%%%%%%%%%%%%%%%%%%%%%%%%%%%%%%%%%%%%%%%%
\section{Introduction}
%%%%%%%%%%%%%%%%%%%%%%%%%%%%%%%%%%%%%%%%%%%%%%%%%%%%%%%%%%%%%%%%%%%%%%%%%%%%%%%%%%%%
% 
Graphics Processing Units (GPUs) came a long way towards desk-side
supercomputers. Despite the fate of other accelerators, they continue
to thrive. Coming originally from the computer gaming sector, NVIDIA
GPUs are now also marketed in a separate line called ``Tesla'', which
meets the requirements of computing centers and HPC cluster
operators. We evaluate the potential of the available NVIDIA CUDA GPU
generations, namely G80, GT200 and GF100 (Fermi), for the kernel of a
flow solver based on the lattice Boltzmann method (LBM) introduced in
section \ref{sec:lbm}.
Early GPU adoptions of the LBM \cite{Li.2003} were based on graphics
APIs, which was a very tedious task. More recent research is focused
on performance \cite{Tolke.2008,Obrecht.2011} as well as more complex
phenomena \cite{Obrecht:ParCFD11} than plain channel flow.
This should give insight into peak performance attainable with
contemporary hardware and show the sustainability of GPGPU
computing. The risk of one single hardware vendor is indisputable for
all applications and certainly limits sustainability in the long
term. In contrast to CUDA, the Open Compute Language (OpenCL) is
applicable to NVIDIA and AMD GPUs as well as CPUs and has gained
interest as a unified programming paradigm in HPC. In section
\ref{sec:GPU} we discuss general features of NVIDIA and AMD GPUs and
suitable parallel programming paradigms and perform STREAM benchmarks.
After that we briefly introduce the WaLBerla framework in section
\ref{sec:walberla}. WaLBerla is a highly parallel software package
for e.g. fluid simulation based on LBM, where we have integrated our
optimized LBM GPU kernels as explained in section \ref{sec:impl}.  In
section \ref{sec:performance} we evaluate the OpenCL performance on
different GPUs in comparison to our CUDA kernel. This gives a
comparison between maximum performance and a maximum variety of
hardware platforms.

%%%%%%%%%%%%%%%%%%%%%%%%%%%%%%%%%%%%%%%%%%%%%%%%%%%%%%%%%%%%%%%%%%%%%%%%%%%%%%%%%%%%
\section{The lattice Boltzmann method} \label{sec:lbm}
%%%%%%%%%%%%%%%%%%%%%%%%%%%%%%%%%%%%%%%%%%%%%%%%%%%%%%%%%%%%%%%%%%%%%%%%%%%%%%%%%%%%
The lattice Boltzmann method (LBM) is a versatile approach to solve
incompressible flows based on a simplified gas-kinetic description of
the Boltzmann
equation~\cite{lba:wolf-gladrow:2000,lba:succi:2001b,lba:chen:1998,lba:qian:1992}.
The LBM operates on a uniform grid of lattice nodes, which are updated
with nearest neighbor information in every time step.

Based on a velocity discrete Boltzmann equation with an appropriate
collision term, e.g. the BGK collision approximation, the LBM
formulates as the following evolution equation:
\begin{minipage}[b]{0.98\columnwidth}
\begin{eqnarray}
  f_i(\vec{x}+\vec{e}_i\delta t,\,t+\delta t) &=& f_i(\vec{x},\,t) - \\
 &&      -\frac{1}{\tau}\left[f_i(\vec{x},\,t) -
                            f_i^\mathrm{eq}(\rho, \vec{u}) \right]\nonumber \\
&& \qquad i=0 \ldots N. \nonumber                          
\label{eq:Boltzmann}
\end{eqnarray}
\end{minipage}
Particle distribution functions $f_i$ represent the probability of
particles at position $\vec{x}$ and time step $t$ with the velocity
$\vec{e}_i$.  $f_{i}^{\mbox{\footnotesize eq}}$ is an approximation of
the Maxwell-Boltzmann equilibrium distribution function at low Mach
numbers and is solely based on the first moments of the particle
distribution function (PDF), i.e. macroscopic fluid density $\rho$ and
macroscopic fluid velocity $\vec{u}$.  The shape of the numerical grid
and the discrete velocity vectors $\vec{e}_i$ are derived from the
discretization applied, i.e. the D3Q19 model~\cite{lba:qian:1992},
which uses 19 discrete velocities in 3-D.

To advance one timestep the following steps need to be taken for each
cell of the domain:

\begin{itemize}
\item Compute the local macroscopic flow quantities $\rho$ and
  $\vec{u}$ from the distribution functions, $\rho = \sum_{i=0}^N f_i$
  and $\vec{u} = \sum_{i=0}^N f_i \vec{e}_i$. 
%\frac{1}{\rho}
\item Calculate the equilibrium distribution $f_i^\mathrm{eq}$ from
  the macroscopic flow quantities (see~\cite{lba:qian:1992} for the
  equation and parameters) and execute the ``collision'' (relaxation)
  process,
  %%% \begin{equation}
    $f_i^* (\vec{x},\,t^*) = f_i(\vec{x},\,t) 
       -\frac{1}{\tau}\left[f_i(\vec{x},\,t) -
                            f_i^\mathrm{eq}(\rho, \vec{u}) \right]$,
  %%% \end{equation}
  where the superscript ``*'' denotes the
  post-collision state (``collide step'').
\item Propagate the $i=0 \ldots N$ post-collision states $f_i^*
  (\vec{x},\,t^*)$ to the appropriate neighboring cells according to
  the direction of $\vec{e}_i$, resulting in $f_i
  (\vec{x}+\vec{e}_i\delta t,\,t+\delta t)$, i.e., the values of the
  next timestep (``stream step'').
\end{itemize}

For arbitrary domains with cells marked as non fluid, i.e. obstacles,
a fourth step must be applied to handle the boundaries, the so-called
``bounce-back'' rule \cite{lba:ziegler:1993}.  This step can be
combined with the third step, the streaming step, as the distributions
are just advected differently. The first two steps can be combined as
the collide step and represent the computationally intensive
part. Usually, implementations choose a two lattice strategy for the
update, which eliminates all data dependencies between last and
current time step. Note that this doubles memory consumption but
simplifies the implementation.

%%%%%%%%%%%%%%%%%%%%%%%%%%%%%%%%%%%%%%%%%%%%%%%%%%%%%%%%%%%%%%%%%%%%%%%%%%%%%%%%%%%%

%%%%%%%%%%%%%%%%%%%%%%%%%%%%%%%%%%%%%%%%%%%%%%%%%%%%%%%%%%%%%%%%%%%%%%%%%%%%%%%%%%%%
\section{GPGPUs and their programming paradigms} \label{sec:GPU}
%%%%%%%%%%%%%%%%%%%%%%%%%%%%%%%%%%%%%%%%%%%%%%%%%%%%%%%%%%%%%%%%%%%%%%%%%%%%%%%%%%%%
\paragraph{Test environment}
All tests have been performed with the CUDA 4.0 Toolkit and the Intel
C Compiler Version 12.0 using the highest optimization level, while
maintaining numerically correct results.

The Intel Westmere EP platform used here accommodates two sockets,
each holding a 6-core processor (Intel Xeon X5650) running at 2.66 GHz
(max. turbo frequency 3.06 GHz) and having 12 MB L3 cache. Each
processor chip forms a ccNUMA (cache coherent non-uniform memory
access) locality domain (LD) since it operates on a single on-chip
memory controller.

The important details of all evaluated GPUs can be seen in
Tab.~\ref{tab:Major-GPU-MP-revisions} and the following section
describes the NVIDIA GPU hardware with the necessary details.

\paragraph{Global memory} The global memory serves as the random
access memory well known from CPUs and is today an order of magnitude
smaller than on most CPU based servers, i.e. $3$ to $6$ GB on the
current Tesla generation (C2050/C2070). The global GPU memory delivers
more than two times the bandwidth, i.e. about $91$~GB/s sustained with
ECC enabled ( $144$~GB/s w/o ECC according to spec sheet) in
comparison to a current standard Intel Westmere based dual socket
server.  ECC stands for Error Checking and Correction and detects and
corrects single bit memory errors. Note that only NVIDIA Tesla cards
starting with the GF100/Fermi generation have ECC protection.  AMD
offers up to $2$~GB of memory on the HD~6970 GPU. An unvectorized
STREAM copy sustains a memory bandwidth of $110$~GB/s, the vectorized
implementation sustains $136$~GB/s ($176$~GB/s w/o ECC according to
spec sheet).

\paragraph{Multiprocessor}
A NVIDIA multiprocessor (MP) consists of several cores which are
driven in a single instruction multiple data (SIMD) manner. Due to the
threading model on the GPU this is also called single instruction
multiple thread (SIMT) model.  In general all cores in a SIMD unit are
controlled by a simple instruction scheduler and have to do exactly
the same instructions, i.e. addition, subtraction, multiplication or
division, at the same time but on different data. Furthermore, the
instruction scheduler cannot alter the sequence of computations
predetermined by the compiler in this strict in-order architecture. In
contrast to CPU hardware there is no large explicit cache on the
GPUs. Newer generations have up to 768~KB of cache. The shared memory
on a GPU is a distinctive feature. In terms of access latency and
transfer speed it is comparable to registers, but it can be accessed
explicitly, similar to the local memory on Cell Broadband Engine
\cite{zcell:ibm:arch:2007}. Hence, there will be no automatic caching
but the implementation has to manage data copying to and from the
shared memory explicitly. On the one hand this gives more
opportunities for optimizations. On the other hand an automatic cache
would give the same improvement, in most cases. For static data, there
is also a memory feature called ``constant memory'', which is read
only but cached. The distinct differences between major multiprocessor
generations can be seen in Tab.~\ref{tab:Major-GPU-MP-revisions}.

\paragraph{SIMD block}

The AMD equivalent to a MP is called SIMD block and consists of
several stream processors. Each stream processor has four units to
process integer and single precision (SP) floating point operations at
the same time, programmed by a VLIW (very long instruction word). In
case of double precision (DP), all these units are utilized
together. So a direct comparison of floating point peak performance
between NVIDIA and AMD depends on the choice of precision
\cite{hpc:AMD6970:2011}. Further specifications like cache
architecture of the AMD hardware are undisclosed and not the focus of
this paper.

\begin{table*}
\centering
\begin{tabular}{|l|l|l|l|l|} \hline
GPU Board& 8800 GTX  & Tesla C1060  & Tesla C2070   &AMD 6970  \\   \hline 
Architecture & G80  & GT200  & GF100   &Cayman  \\   \hline 
 Multiprocessors &16 &30 &14 &24   \\   \hline
Total Cores  &128 &240 &448 &1536    \\   \hline
Compute Capability &1.0 & 1.3  &2.0 & -- \\ \hline
Memory [GB] & 0.768  & 4  & 6   &2  \\   \hline
Memory BW [GB/s] & 86.4 & 102.4 & 140  & 176 \\ \hline  
Peak FP SP [GFLOP/s] & 518   & 933   & 1030    &2700   \\   \hline 
Peak FP DP [GFLOP/s] & N/A   & 77   & 515    &683   \\   \hline 
\hline
\multicolumn{5}{|l|}{Multiprocessor:}  \\ \hline
Cores & 8 & 8  & 32 & 16 \\ \hline  
Maximum \# thread &1024 &1024 & 1536 & 256   \\   \hline
%e
L2 Cache [kB]  &$*$   &$*$   & 768  &$*$  \\ \hline 
Shared memory [kB]  & 16  & 16  & 16/48  &$*$    \\ \hline 
Texture Cache [kB]  & 16  & 16  & 48/16  &$*$    \\ \hline 
Register  & 8192 & 16384 & 32768 &$*$    \\ \hline 
\end{tabular}
\caption{ Major NVIDIA GPGPU multiprocessor revisions with documented
  features. The size of shared memory and texture cache are
  adjustable on the GF100. AMD specifications are provided where applicable.($*$ Feature not documented.)
}
\label{tab:Major-GPU-MP-revisions}
\end{table*}

\paragraph{Host to device interface}
The host to device interface is currently PCIe Gen2 which has a
maximum transfer bandwidth of 16 GB/s (according to the
specifications) between host and device. Although transfers over this
bus are not covered in this work, it is well known that the PCIe bus
represents one of the major bottlenecks in GPU computing
\cite{gpu:habich:ades}.
 
\subsection{CUDA software environment}
The compute unified device architecture (CUDA) was the first
approach to GPGPU computing without the hassle to mimic arithmetic
operations as graphical operations. Each task is split up into a host
(CPU) and device (GPU) part, the latter is also called kernel. The
underlying programming language is C, which got extended by several
keywords. The NVCC (NVIDIA C-Compiler) understands these extensions
and compiles a GPU capable executable. NVIDIA C++ extensions are not
discussed in this work. 

Explicit kernel calls are necessary to invoke GPU
computations. Required parameters, the number of blocks and the number
of threads, are specified in this call. The finest level of
parallelism is a GPU thread. Each thread is assigned to one specific
block. The execution of threads is scheduled in packages called
``warps'', each comprising 32 threads.
For the current GF100 32 threads start simultaneously. As a general
guideline the minimum feasible number of threads to schedule per block
is 32 and larger thread numbers should be a multiple of that. The
programmer has to distribute the work dependent on the threadID and
blockID of the current thread. No automatic work sharing construct
(e.g., for loops as known from OpenMP \cite{hpc:OpenMP:2011} ) is
implemented.
On the G80 architecture one single blocking kernel at a time was
issued and the runtime system returned back to the CPU process after
the kernel was finished. Starting with CUDA 3.2 and compute capability
2.0 one can issue multiple non-blocking kernels in parallel, which
actually get executed in parallel by utilizing different so called GPU
streams. Inside a stream, kernel calls and memory copies are kept in
the order of their submissions. Tasks of different streams can share
the same GPU and overlap work. The runtime system decides on how to
best place work items. This gives opportunity for algorithms with low
parallelism to perform concurrently with other tasks or to overlap
communication and CPU to GPU data copy. Note that overlapping most
likely does not happen for kernels, where a single kernel already
utilizes all GPU resources. For more details see
\cite{gpu:nvidiaCudaProgrammingGuide4.0:2011}.

\subsection{OpenCL}
The Open Compute Language (OpenCL)~\cite{hpc:KHRONOS:OpenCL:2011}
(maintained by the a consortium of different hard- and software
vendors, e.g. NVIDIA and AMD called KHRONOS GROUP) is currently the
most comprehensive approach to simplify and generalize access to
compute capabilities offered by GPUs, CPUs and other present and
future accelerators. Like CUDA, OpenCL programs can be written in C
and C++ like syntax. Although the basic terminology is different,
e.g. a CUDA block is a Workgroup and a CUDA thread is a Workitem, the
guidelines for programming OpenCL and CUDA on the same hardware are
identical or at least very similar.

In this paper we focus on the existing CUDA architecture and adhere to
the CUDA nomenclature and definitions.

\subsection{Massively parallel threaded execution}
Access to the global device memory has much higher hardware latencies
than known from main memory access on CPUs. Mainly the absence of
mechanisms to hide latencies, i.e.  caches, hardware prefetching
algorithms, rescheduling of the compiled instructions (out of order
scheduling on CPUs) are the reasons for this on GPUs.
Massive thread oversubscription is the way to hide the latency on
GPUs.
Threads that are waiting for memory IO, basically a bubble in the
execution pipeline, are interleaved with those which have work
scheduled.  The instruction scheduler unit can switch between warps of
blocks very efficiently and instantly, thus, filling these bubbles in
the execution pipeline and data pipeline.  This architectural feature
compensates the high latency of the memory system.
Currently a two socket server has up to 24 cores available. On an
NVIDIA GPU we now have 448 cores, so in general at least 448 threads
are necessary to utilize all cores.  On CPUs one or at most two
(Simultaneous Multi Threading) high demand processes or threads are
usually scheduled per core to minimize costly process or thread
switching and migrations. In contrast to that much more threads can be
issued on a GPU per hardware core without a drawback in
performance. In general a given problem has to be decomposed into at
least $1536*14=21504$ independent threads to reach the maximum number
of concurrently scheduled threads.
%

%\vspace{5mm}
\begin{figure}[htbp]
  \centering
  \includegraphics[width=0.99\columnwidth,clip=on]{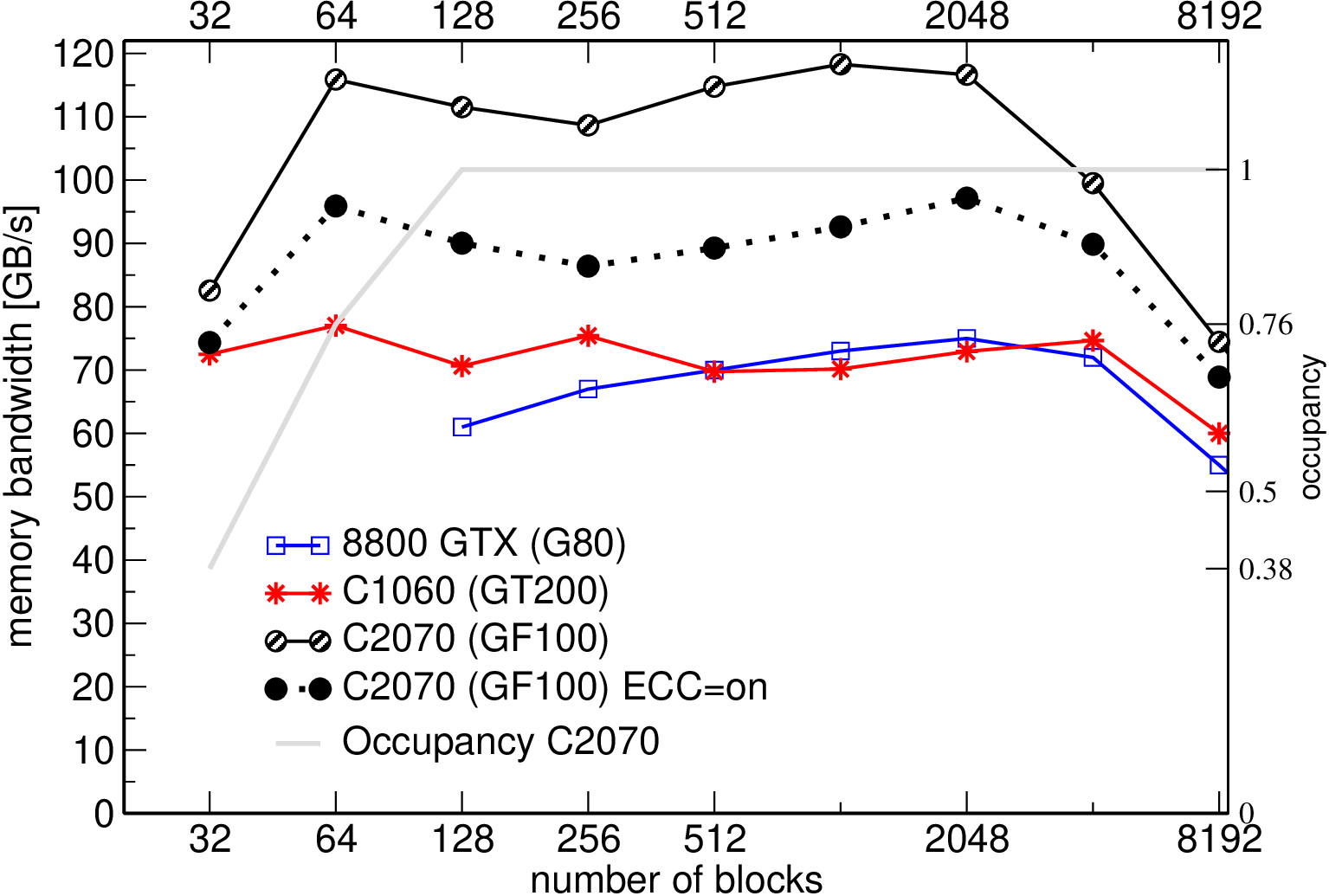}
  \caption{\label{chart:VectorCopyG80-GT200-GT300} Overview of
    performance of STREAM copy measurements on different NVIDIA GPU
    generations for a vector size of $2^{20}$ and 256 threads per CUDA
    threadblock.  }
\end{figure}

\vspace{5mm}

The fast scheduling of different threads and warps comes at a
price. All resource constraints of threads on the MP have to be
satisfied right from the start till one block is completely
finished. In particular all active threads share the available
registers and shared memory on one MP (see
Tab.~\ref{tab:Register-per-Thread}).  Only $20$ registers are
accessible for each thread at full occupancy to execute the
kernel. Occupancy is defined by the number of actual concurrently
running threads divided by the maximum schedulable number of threads.
\begin{table}
\centering
\begin{tabular}{|l|c|c|c|c|c|} \hline
Threads & 1536   & 1024  & 512 & 256 & 64    \\   \hline 
%\multicolumn{6}{|l|}{GF100} \\   \hline 
GF100  & 20   & 30  & 62 & 125 & 500    \\   \hline 
%\multicolumn{6}{|l|}{GT200} \\   \hline 
GT200  & --   & 64  & 32 & 64 & 256    \\   \hline 
%\multicolumn{6}{|l|}{G80} \\    \hline 
G80  & --   & --  & 16 & 32 & 128    \\   \hline 
\end{tabular}
\caption{
Number of registers available per Thread.
}

\label{tab:Register-per-Thread}
\end{table}

Increasing kernel concurrency is not sufficient to ensure high
performance for memory access. In addition one has to keep in mind
coalesced loads and stores and implement the memory access
accordingly. Basically threads executed concurrently on a MP need to
access data with high spatial locality to get memory requests
bundled. This way, latency has to be paid only once for each
contiguous memory block. The first G80 based architecture required a
strict mapping between thread index and data index. Each thread had to
access the element with its thread index starting at an 128 byte
aligned address. GT200 introduced caches
\cite{Volkov:2008:BGT:1413370.1413402}, which relaxed the thread to
data mapping substantially. High memory bandwidth can be attained as
long as the concurrent data accesses of all active threads have a high
spatial data locality. Still an improvement can be seen from correct
coalescing. The latest GF100 architecture improved scattered data
access even more and only marginal improvements can be seen by correct
coalescing.

\subsection{Case study STREAM benchmarks}
The attainable memory bandwidth of GPUs and CPUs can be determined
with the STREAM benchmarks \cite{hpc:StreamBench:2011}.  The STREAM
copy \texttt{C=A}, where \texttt{A} and \texttt{C} are large vectors,
mimics the load to store ratio of the LBM method. 

\paragraph{GPU}
Measurements for all NVIDIA GPGPU generations can be seen in
Fig.~\ref{chart:VectorCopyG80-GT200-GT300}. Note that 8800~GTX is a
consumer card. The Tesla series cards are clocked lower than consumer
cards of the same generation. The strong scaling of a vector with
$2^{20}$ elements with different numbers of blocks and 256 threads per
block shows that the scheduling system works very efficiently. We see
a lower performance for only 32 blocks as the MPs run only with
suboptimal occupancy on C2070. Once occupancy reaches at least $0.5$
we see peak memory performance. At 8192 blocks we observe a
performance reduction, which correlates to $8192 *256 =2* 2^{20} $. So
half of all threads run empty, which of course impacts
performance. The behavior is consistent over all architectures.  For
larger problem sizes, the performance breakdown will shift
accordingly, and occur at larger block sizes.
The exact implementation of ECC on NVIDIA cards is not publicly
available. In general ECC~\cite{hpc:Wiki:ECC} is implemented in such a
way that for 8~Bytes one additional Byte is stored with redundancy
information, i.e. $12.5$~\% overhead. This is supported by our STREAM
measurements as ECC has 10~\% to 18~\% less user data bandwidth. Note
that ECC data reduces the total amount of available space for user
data on the GPU (e.g. $5.25$~GB instead of $6$~GB, which corresponds
to $12.5$~\% redundancy). Furthermore, ECC checksums are calculated in
the same arithmetic units as user data and transfered over the same
memory subsystem. So immanently, any change of the ECC state changes
the data alignment and access pattern as well.

\paragraph{CPU}
The OpenMP parallelized STREAM benchmark performance is shown in
Tab.~\ref{tab:stream} and is about $40$\% of the GPU bandwidth with
ECC. Note that ECC on this server cannot be disabled.

\begin{table}[htbp]

\center
\begin{tabular}[b]{l|ccc}
                   & 1 core      & 1 NUMA LD       & 1 node \\ \hline

Intel X5650           & 10.01     & 14.08        & 26.8   \\ % 41/255
\small measured [GB/s] &&&\\
Intel X5650 & 15.01    & 21.12        & 40.2   \\ % 41/255
\small actual [GB/s]  &&&\\
\end{tabular}
%
%\end{minipage}
%\begin{minipage}[b]{0.33\textwidth}
\caption{\label{tab:stream} STREAM copy performance in GB/s. The actual bandwidth accounts for the write allocate.}
%\end{minipage}
\end{table}

\section{Performance model}
For GPU and CPU implementations of the LBM the first major performance
concern is the memory bandwidth, 
%since the vast computational
%capabilities of GPUs and CPUs cannot be utilized due to the
owing to the computational balance of the LBM of roughly $1.5$~Bytes
per FLOP in DP \cite{2006:Wellein:CompFluids}. In contrast to that an
NVIDIA C2070 has a system balance of 0.27 Bytes/FLOP and an Intel Xeon X5650 dual
socket node has 0.16 Bytes/FLOP.

Hence, based on the attainable memory bandwidth one can establish a
basic performance model for CPUs and GPUs. For the LBM we have to
consider $19$ distributions being loaded from and stored to memory.
The bytes transferred for each LBM lattice cell update can
be determined by:
\[n_{bytes} = n_{stencil} \cdot (n_{loads}+n_{store}) \cdot s_{PDF},\]
where $n_{stencil}$ is the size of the LBM stencil, $n_{loads}$ and
$n_{stores}$ are the number of loads and stores and $s_{PDF}$ is the
size in bytes of a single PDF variable.
A peculiarity of cache based architectures like CPUs is that a store
to a memory location requires this data to be in the cache. Otherwise
a ``write allocate'' is necessary to fetch the
data to the cache. A two lattice implementation leads to 3 total
memory transfers on CPUs and to a total of $228$~Bytes using single
precision  and $456$~Bytes using DP. On the GPU a
total of $152$~Bytes using single precision  and $304$~Bytes using
double precision is transferred for a single lattice cell update.
The performance of LBM codes is usually given in terms of {\em million
  fluid lattice cell updates per second} (MFLUP/s) instead of GFLOP/s
as it is not a feasible performance metric for LBM.
The maximum sustainable performance can be derived by dividing the
bandwidth obtained from the STREAM benchmarks in Tab.~\ref{tab:stream}
and Fig.~\ref{chart:VectorCopyG80-GT200-GT300} by the number of Bytes
needed for a lattice update.

Tab.~\ref{tab:PerformanceModel} gives the upper performance estimates
based on the attainable STREAM bandwidth for the lattice Boltzmann
method on the presented architectures.
%
%
%\end{eqnarray}
\begin{table}[htbp]
%\begin{minipage}[b]{0.66\textwidth}
\center
\begin{tabular}[b]{l|rrr}
                & SP    & DP \\ \hline
Intel X5650 node  & 176   & 88 \\ 
C2070           &788    &394 \\
C2070 ECC       &624    &312 \\
C1060           & 512   &256 \\
G80             &492    & N/A \\
\end{tabular}
%\end{minipage}
%\begin{minipage}[b]{0.33\textwidth}
\caption{\label{tab:PerformanceModel} Performance model for CPU and
  GPUs in MFLUP/s based on the STREAM bandwidth measurements.}
%\end{minipage}
\end{table}

\section{The WaLBerla framework} \label{sec:walberla} 
WaLBerla \cite{hpc:Feichtinger:2011} is a massively parallel
multiphysics software framework that is originally centered around the
LBM, but whose applicability is not limited to this algorithm.
Its main design goals are to provide excellent application performance
across a wide range of computing platforms and the easy integration of
new functionality.  In this context additional functionality can
either extend the framework for new simulation tasks, or optimize
existing algorithms by adding special-purpose hardware-dependent
kernels or new concepts such as load balancing strategies.
Despite the overhead of a large framework, WaLBerla proved to be as
fast as standalone kernels~\cite{Feichtinger:2010:ParCo}.
Various complex simulation tasks have already been incorporated into
WaLBerla.  Amongst others, free-surface
flows~\cite{2010:Donath:CompFlu} particulate flows for several million
volumetric particles~\cite{2010:Goetz:SC} on up to around $3000000$
cores have been included.

%%%%%%%%%%%%%%%%%%%%%%%%%%%%%%%%%%%%%%%%%%%%%%%%%%%%%%%%%%%%%%%%%%%%%%%%%%%%%%%%%%%%
\section{Lattice Boltzmann method implementation} \label{sec:impl}
%%%%%%%%%%%%%%%%%%%%%%%%%%%%%%%%%%%%%%%%%%%%%%%%%%%%%%%%%%%%%%%%%%%%%%%%%%%%%%%%%%%%

\paragraph{GPU} 
The application provides sufficient parallelism on the GPU as a single
thread computes the update of a single lattice node. The
stream--collide sequence was chosen in contrast to
\cite{gpu:habich:ades}. Instead of the local PDFs the collision is
computed based on the PDFs of the surrounding lattice cells. The
updated PDFs are then stored to the local lattice cell. Hence, the
algorithm loads scattered data from the memory, which however, seems
to have only very low impact on coalescing and
performance. Furthermore, the available data caches support the
scattered memory load if the data locality of all threads on the MP is
high. The store to main memory happens only at the local cell index
and is perfectly coalescable.  Note that two grids ensure that no data
dependency is violated.

Since, the local stores are aligned it is obsolete to reroute write
access through shared memory. Loading coalesced data manually and
storing it temporarily in shared memory for later access proved not to
be advantageous.

As mentioned above a minimal occupancy of $0.5$ is needed to achieve
high memory bandwidths. With $2 \times 19$ distribution functions this
limit is rather low since the initial implementation led to $100$
registers being used in the kernel. Hence, techniques like arithmetic
optimizations and minimizing temporary variables have to be
employed. Again, storing intermediate data temporarily in the shared
memory was not advantageous.

An optimized index operator decreased the register usage down to 32
registers. This technique can be applied to any algorithm working on
multidimensional data.

In order to load data perfectly coalesced it has to be aligned to 128
Byte. However, domainsizes are often not obeying this alignment
constraint and padding cells are introduced after each stripe of
elements. The padding size has to be chosen in such a way, that the
next simulation cell starts at an aligned address. Furthermore, the
algorithm has to be altered to not compute on the padding elements.

\begin{figure}[htbp]
\centering
                \includegraphics[width=0.99\columnwidth,clip=on]{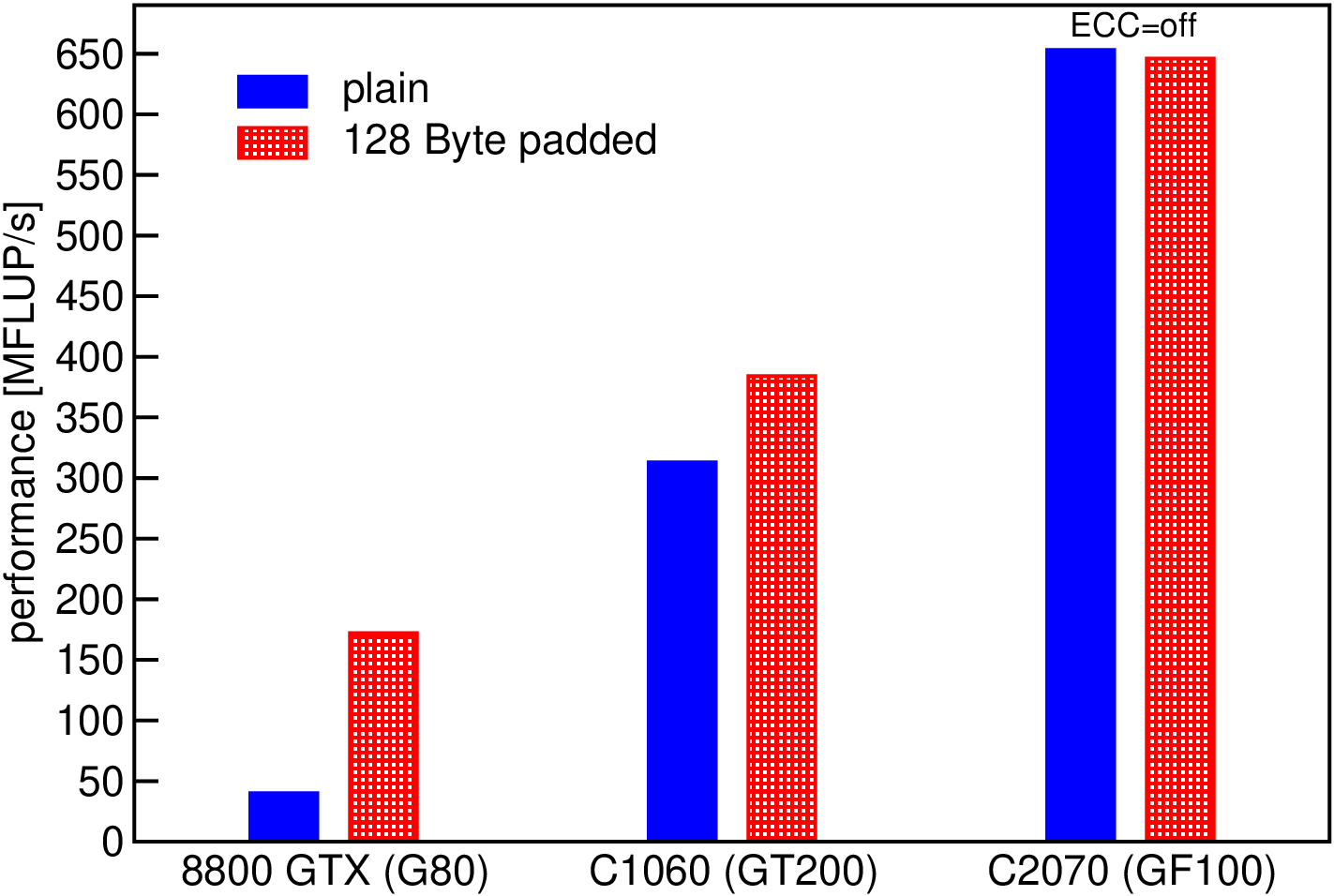}

                \caption{LBM Performance on different NVIDIA GPGPU
                  generations in single precision with a domainsize of
                  $200^3$ ($158^3$ on 8800~GTX owing to limited global
                  memory). Padding introduces overhead cells after
                  each stripe to ensure correct
                  alignment.\label{chart:LBMG80-GT200-GF100-sp}}
\end{figure}
\begin{figure}[htbp]
\centering
                \includegraphics[width=0.99\columnwidth,clip=on]{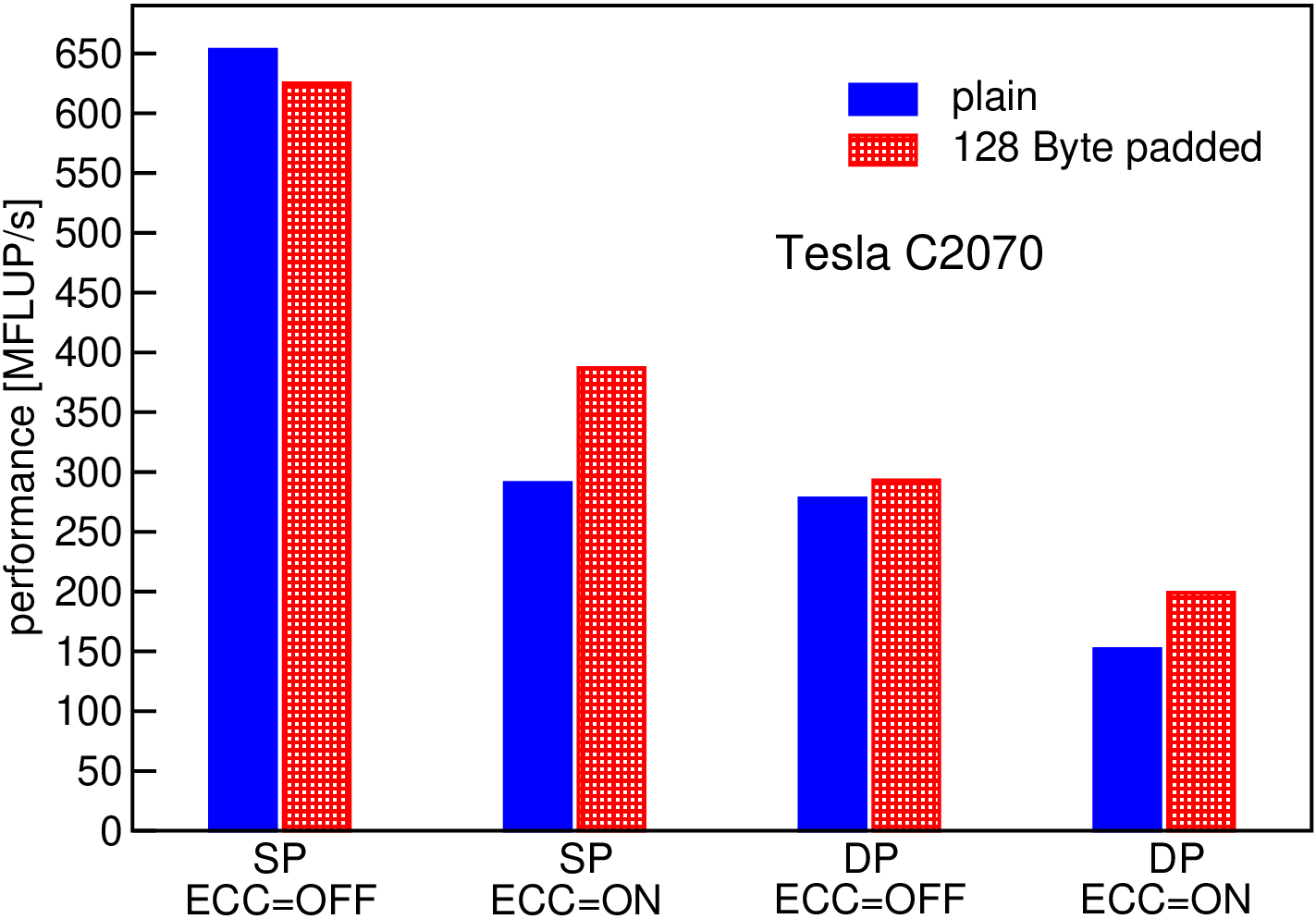}

                \caption{LBM Performance with and without ECC for
                  single and double precision on C2070 (GF100) with a
                  domainsize of $200^3$. For ECC enabled setups and
                  simulations in double precision, padding is
                  beneficial. Only for single precision without ECC,
                  padding has a small negative
                  effect.  \label{chart:GF100-sp-vs-dp-vs-ECC-bar}}
\end{figure}

\begin{figure}[htbp]
\centering
                \includegraphics[width=0.99\columnwidth,clip=on]{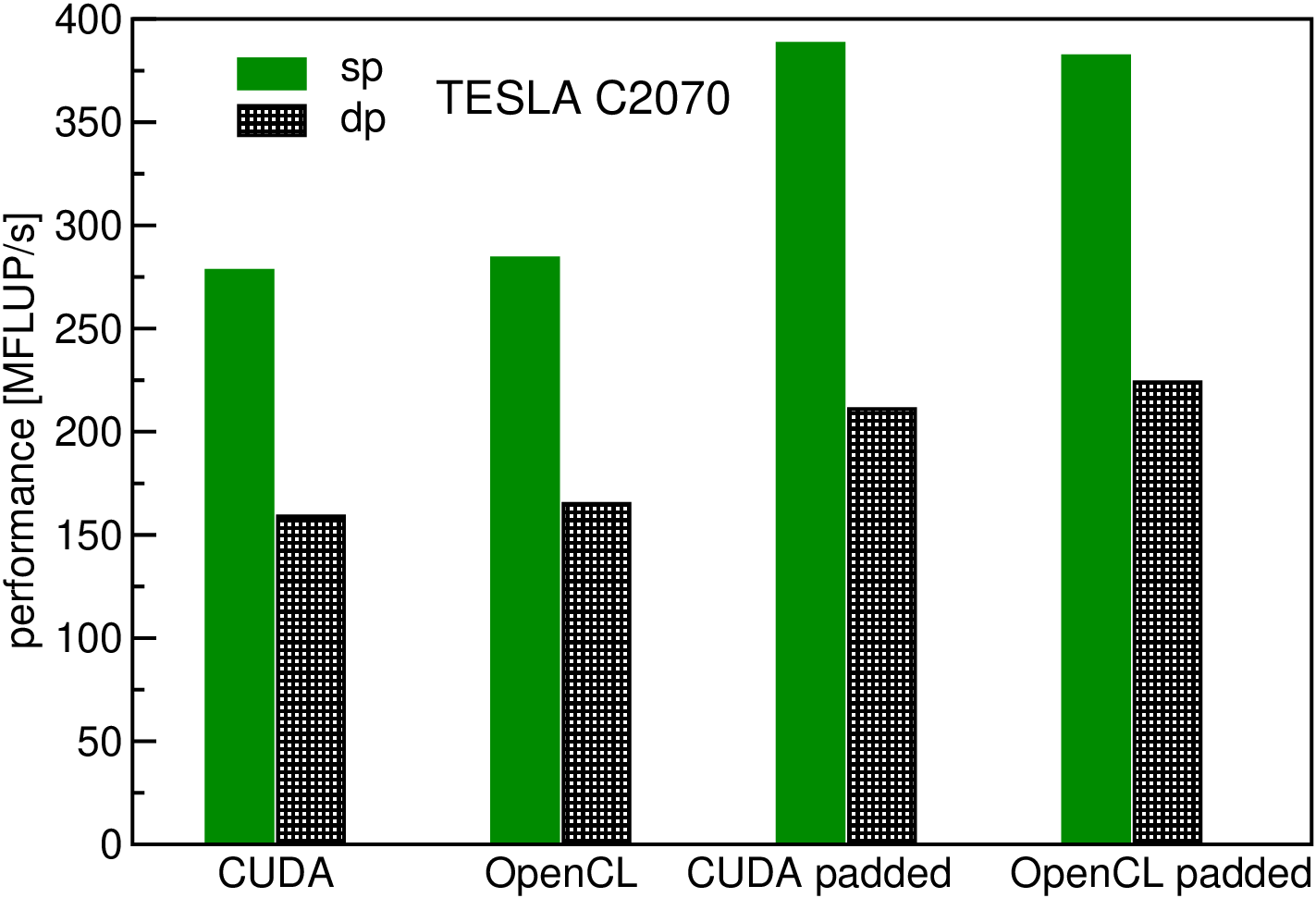}
                \caption{LBM Performance of CUDA and OpenCL
                  implementation on C2070 with ECC with a domain size
                  of $200^3$.\label{chart:CUDA-OCL-bar}}
\end{figure}
\section{Performance results} \label{sec:performance}
The optimized implementation uses one kernel routine for the
stream-collide step, but in contrast to general CPU implementations
\cite{Wellein:ParCFD11}, the treatment of non-fluid cells is done in
separate kernel. A simple lid-driven cavity problem on different cubic
domains was employed to verify the optimized CUDA and OpenCL
implementations.
\paragraph{CUDA}
The performance of our LBM implementation can be seen in
Fig.~\ref{chart:LBMG80-GT200-GF100-sp} for single precision (SP) ( DP
is not supported on the first generation.)  and ECC disabled to allow
for a comparison with older GPU generations.

We see a two fold speedup from 8800~GTX to C1060, although the memory
bandwidth has improved only marginally. The performance gain is mainly
due to doubling the available register space on GT200. Consequently,
occupancy is twice as before and the sustainable memory bandwidth
doubles for our LBM implementation.

Obviously, correct alignment and padded memory access was mandatory on
G80 and favorably on GT200 based GPUs. Padding is no longer relevant
on GF100 based GPUs for single precision calculations if ECC is
disabled. Instead it is counterproductive as additional memory
transfers decrease performance slightly.  Overall improvement from
C1060 to C2070 is about 40~\% which corresponds to the increase in
memory bandwidth.

For double precision and ECC enabled the influence of padding is as
anticipated on C2070 as well.
Fig.~\ref{chart:GF100-sp-vs-dp-vs-ECC-bar} shows that ECC enabled
setups benefit from appropriate padding. Note, that already 16~Byte
padded access to memory increases performance but 128~Byte padding
gives best performance.

The large performance gap between ECC and non-ECC simulations cannot
be explained solely by the additional data transfers (see above).
This issue here is currently under investigation.
%  \vspace{5mm}
\begin{figure*}[htbp]
\centering
\begin{minipage}[t]{1\columnwidth}
                \includegraphics[width=0.99\columnwidth,clip=on]{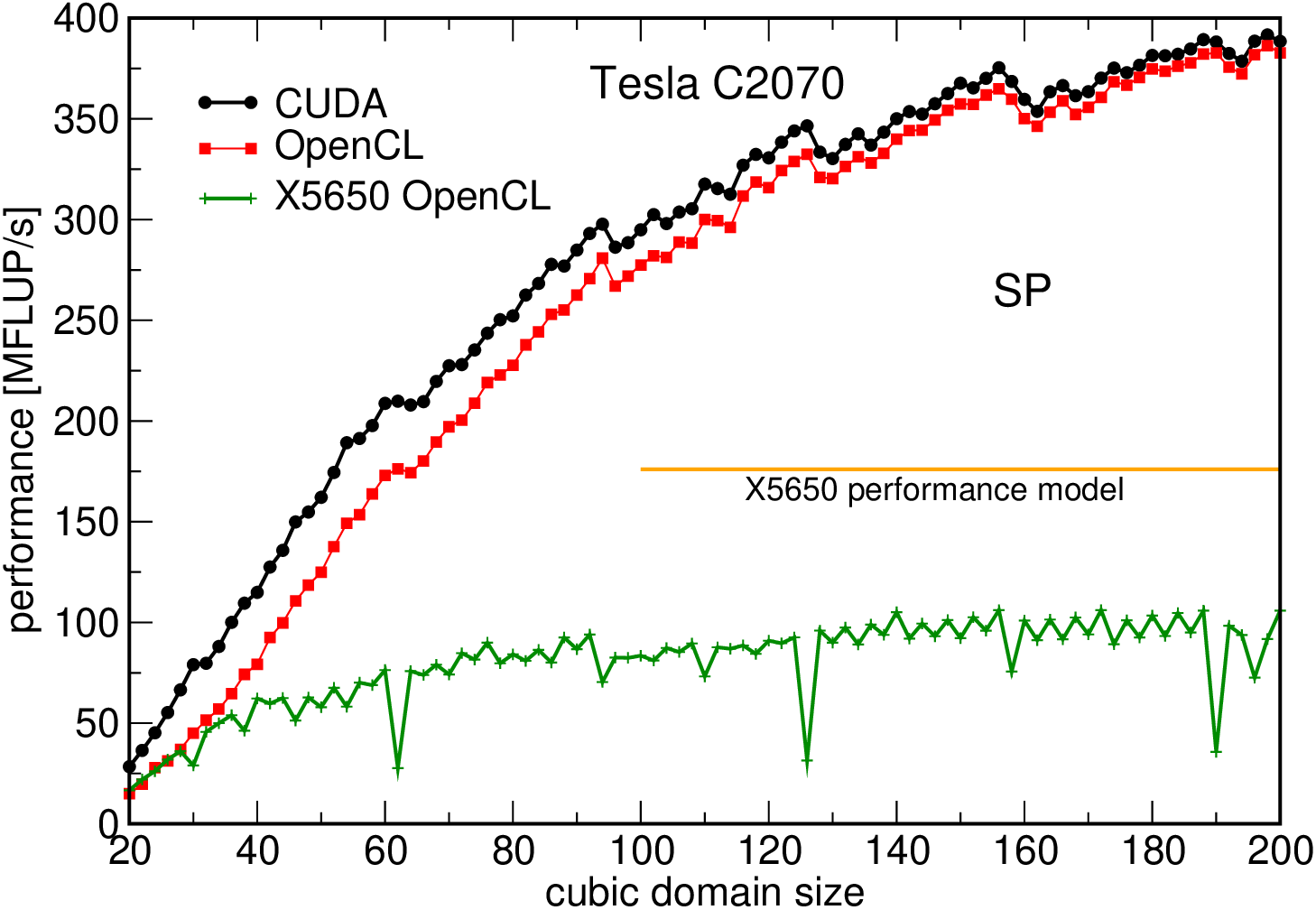}

\end{minipage}
\hfill
\vspace{5mm}
\begin{minipage}[t]{1\columnwidth}
                \includegraphics[width=0.99\columnwidth,clip=on]{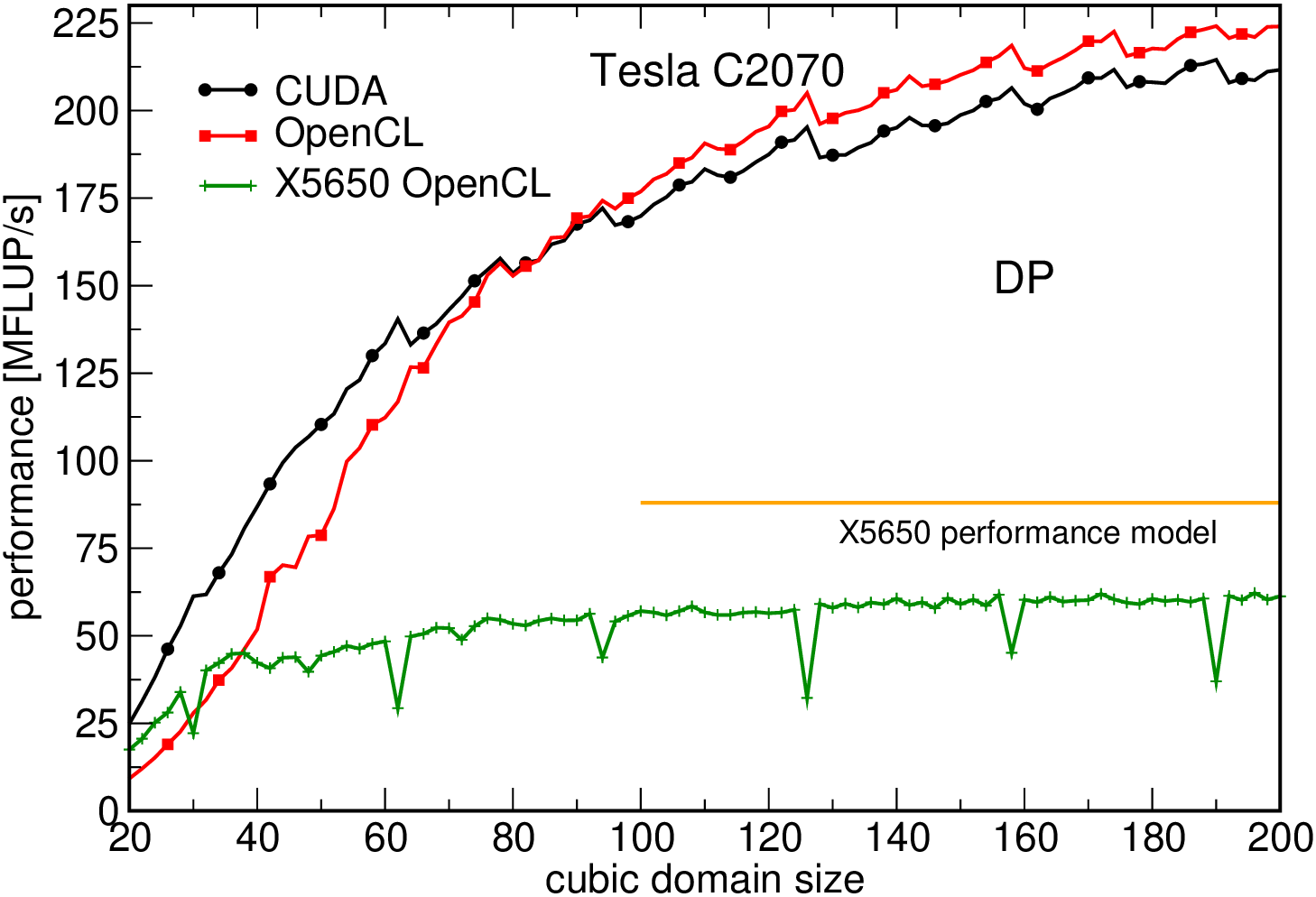}

\end{minipage}
                \caption{Performance comparison of CUDA and OpenCL on
                  C2070 with padding in SP (upper)  and DP (lower) with ECC. Furthermore,
                  measurements with the same OpenCL implementation on
                  the Intel Xeon X5650 CPU are shown sustaining about $60$~\% of the
                  performance  model in SP and $71$~\% in DP.
                  \label{chart:Walcore-domainsweep-C2070-Cuda-OCL-ECC-sp-padded}
                  \label{chart:Walcore-domainsweep-C2070-Cuda-OCL-ECC-dp-padded}
                }
 
\end{figure*}

\paragraph{OpenCL GPU}
Next we compare the performance of our CUDA implementation to an
OpenCL implementation. The CUDA kernels needed minimal
adjustments to work under OpenCL. Basically the calculation of the
local index by taking CUDA threadID and blockID were replaced by
OpenCL's localID and groupID. The remaining kernel adjustments affect
only function declarations. The performance of OpenCL in comparison to
CUDA can be seen in Fig.~\ref{chart:CUDA-OCL-bar}.  Qualitatively
speaking there is no difference in ECC enabled setups for single and
double precision. Padding has the same positive influence as for CUDA.
%\hfill
\vspace{5mm}
\begin{figure}[htbp]
\centering
                \includegraphics[width=0.99\columnwidth,clip=on]{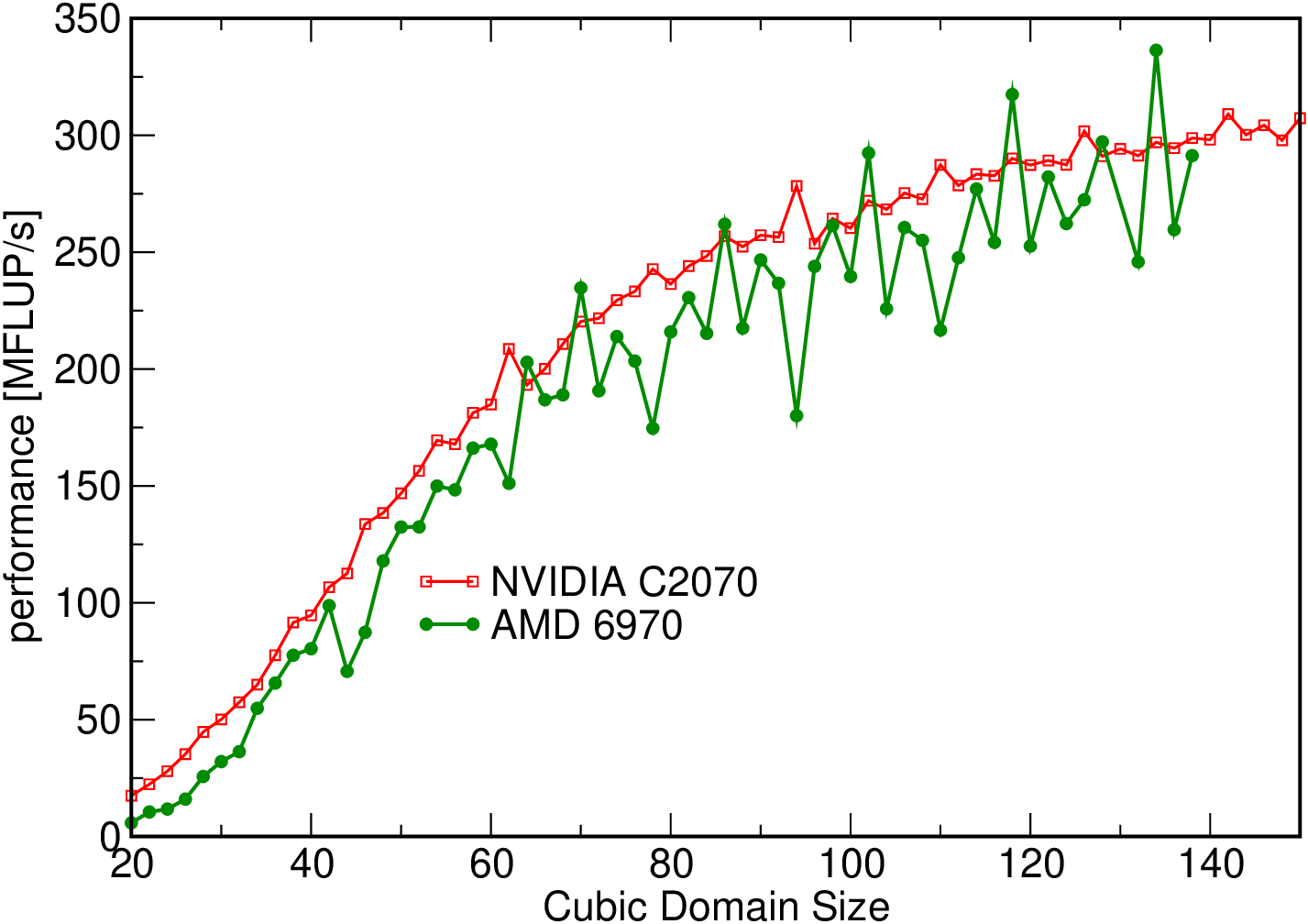}
                \caption{Performance comparison of AMD 6970 and NVIDIA
                  Tesla C2070 with OpenCL in DP without
                  ECC.\label{chart:Walcore-domainsweep-C2070-AMD6970-dp-noecc-padded}}
\end{figure}

A full domain sweep is shown in
Fig.~\ref{chart:Walcore-domainsweep-C2070-Cuda-OCL-ECC-sp-padded} for
SP and DP. In SP OpenCL is on par with CUDA on the GPU. For DP OpenCL
is indeed faster by 6\% for large domains. Both algorithms show the
same overall behavior, although CUDA is faster for domains up to
$80^3$. The small peaks occurring at domain sizes of multiples of 16
are perfectly aligned and no padding needs to be applied.
Benchmarks for the OpenCL enabled AMD HD~6970 can be seen in
Fig.~\ref{chart:Walcore-domainsweep-C2070-AMD6970-dp-noecc-padded},
compared to the results of the NVIDIA C2070.  Overall performance is
comparable to the C2070 performance but more erratic. A possible
reason is the missing cache hierarchy which eases latency hiding on
the C2070. Based on the memory bandwidth of the spec sheet one would
anticipate a $19$\% higher performance on HD~6970. However, our code
was thoroughly optimized and tested for CUDA and NVIDIA GPUs, but not
for OpenCL and AMD based GPUs. These results are very encouraging for
further GPGPU development as OpenCL has caught up with CUDA and
provides access to more versatile hardware and even CPUs.

\paragraph{OpenCL CPU} 
Measurements of OpenCL on the CPU are shown in
Fig.~\ref{chart:Walcore-domainsweep-C2070-Cuda-OCL-ECC-sp-padded} as
well. Comparing the DP CPU performance to the performance model we
obtain $71$~\% of performance on this architecture. So there is still
room for improvement. For SP we do not see twice the performance as in
DP in contrast to our estimates of the performance model in
Tab.~\ref{tab:PerformanceModel}. Obviously, memory bandwidth is not
the issue here. A look at the performance counters for vectorized and
scalar FLOPS with likwid \cite{10.1109/ICPPW.2010.38,hpc:likwid} shows
just scalar operations. Thus, only a fourth of the possible SP FLOPS
of the Intel XEON can be utilized.

\section{Conclusion}
We here presented a highly optimized LBM based kernel for GPGPUs in
CUDA and OpenCL. Since the LBM is mainly memory bandwidth bound and a
single lattice update takes 304 Byte on the GPU and 456 Byte on the
CPU in DP, we initially established an upper performance limit based
on sustainable memory bandwidth by implementing benchmarks for basic
memory operations, the STREAM benchmarks. Up to $110$~GB/s can be
sustained on a NVIDIA C2070 GPU ($95$~GB/s with ECC enabled) and up to
$40$~GB/s on the Intel XEON X5650 host node. The usage of ECC shows a
loss of $10$\% to $18$~\% on the GPU which is close to the limit of
$12$~\% suggested by the mostly undisclosed documentation and the
bandwidth overhead.

The same LBM kernel structure is applied in both CUDA and OpenCL and
leads to $83$~\% of the sustainable memory bandwidth in single
precision without ECC memory.  Enabling ECC leads to a substantial
loss in performance between $30$ and $40$~\%.  To determine the reason
for this discrepancy is the topic of current research. However, using
ECC is inevitably necessary to ensure correct simulation results and
should not be neglected to favor faster execution. Only NVIDIA Tesla
cards currently support ECC memory.  Still the GPU kernels give a
speedup of two in contrast to a full two socket Intel Xeon server with
optimized C intrinsics \cite{Wellein:ParCFD11}.

The implemented OpenCL kernel performs well on both, NVIDIA and AMD
GPUs with the same performance. Furthermore, the performance is on par
with the CUDA measurements.

The OpenCL GPU kernel performs surprisingly well on the CPU, which
reaches $71$~\% of the performance model. These are first promising
results, as no optimizations for the CPU have been applied. A speedup
of less than two from DP to SP and the absence of any packed
instructions leads to the conclusion that the OpenCL compiler is not
capable of vectorizing the code.

%%%%%%%%%%%%%%%%%%%%%%%%%%%%%%%%%%%%%%%%%%%%%%%%%%%%%%%%%%%%%%%%%%%%%%%%%%%%%%%
\subsection*{Acknowledgements}
%%%%%%%%%%%%%%%%%%%%%%%%%%%%%%%%%%%%%%%%%%%%%%%%%%%%%%%%%%%%%%%%%%%%%%%%%%%%%%
This work was partially funded by the ``Bundes\-ministerium f{\"u}r
Bildung und Forschung'' under the \emph{SKALB} project,
no. 01IH08003A, and by the ``Kompetenznetzwerk f{\"u}r
Tech\-nisch\--Wis\-sen\-schaft\-liches Hoch- und
H{\"o}chst\-leistungs\-rechnen in Bayern'' (\emph{KONWIHR}) via
\emph{OMI4papps}. The authors thank Fujitsu for providing the ``Sandy
Bridge'' platform through an early access program.

%%%%%%%%%%%%%%%%%%%%%%%%%%%%%%%%%%%%%%%%%%%%%%%%%%%%%%%%%%%%%%%%%%%%%%



%% References


\bibliographystyle{model3-num-names}
\bibliography{bibfile}

\begin{thebibliography}{26}
\providecommand{\natexlab}[1]{#1}
\providecommand{\url}[1]{\texttt{#1}}
\providecommand{\urlprefix}{URL }
\expandafter\ifx\csname urlstyle\endcsname\relax
  \providecommand{\doi}[1]{doi:\discretionary{}{}{}#1}\else
  \providecommand{\doi}{doi:\discretionary{}{}{}\begingroup
  \urlstyle{rm}\Url}\fi
\providecommand{\eprint}[2][]{\url{#2}}
\providecommand{\BIBand}{and}
\providecommand{\bibinfo}[2]{#2}
\ifx\xfnm\undefined \def\xfnm[#1]{\unskip,\space#1}\fi
%Type = Article
\bibitem[{Li et~al.(2003)Li, Wei and Kaufman}]{Li.2003}
\bibinfo{author}{Li\xfnm[ W.]}, \bibinfo{author}{Wei\xfnm[ X.]},
  \bibinfo{author}{Kaufman\xfnm[ A.]}.
\newblock \bibinfo{title}{Implementing lattice b{o}ltzmann computation on
  graphics hardware}.
\newblock \bibinfo{journal}{The Visual Computer}
  \bibinfo{year}{2003};\bibinfo{volume}{19}:\bibinfo{pages}{444--456}.
\newblock \urlprefix\url{http://dx.doi.org/10.1007/s00371-003-0210-6}.
%Type = Article
\bibitem[{T{\"o}lke and Krafczyk(2008)}]{Tolke.2008}
\bibinfo{author}{T{\"o}lke\xfnm[ J.]}, \bibinfo{author}{Krafczyk\xfnm[ M.]}.
\newblock \bibinfo{title}{Teraflop computing on a desktop pc with gpus for 3d
  cfd}.
\newblock \bibinfo{journal}{International Journal of Computational Fluid
  Dynamics}
  \bibinfo{year}{2008};\bibinfo{volume}{22}(\bibinfo{number}{7}):\bibinfo{page%
s}{443--456}.
%Type = Article
\bibitem[{Obrecht et~al.(2011{\natexlab{a}})Obrecht, Kuznik, Tourancheau and
  Roux}]{Obrecht.2011}
\bibinfo{author}{Obrecht\xfnm[ C.]}, \bibinfo{author}{Kuznik\xfnm[ F.]},
  \bibinfo{author}{Tourancheau\xfnm[ B.]}, \bibinfo{author}{Roux\xfnm[ J.J.]}.
\newblock \bibinfo{title}{A new approach to the lattice boltzmann method for
  graphics processing units}.
\newblock \bibinfo{journal}{Computers {\&} Mathematics with Applications}
  \bibinfo{year}{2011}{\natexlab{a}};\bibinfo{volume}{61}(\bibinfo{number}{12}%
):\bibinfo{pages}{3628--3638}.
%Type = Inproceedings
\bibitem[{Obrecht et~al.(2011{\natexlab{b}})Obrecht, Kuznik, Tourancheau and
  Roux}]{Obrecht:ParCFD11}
\bibinfo{author}{Obrecht\xfnm[ C.]}, \bibinfo{author}{Kuznik\xfnm[ F.]},
  \bibinfo{author}{Tourancheau\xfnm[ B.]}, \bibinfo{author}{Roux\xfnm[ J.J.]}.
\newblock \bibinfo{title}{Multi-gpu implementation of a hybrid thermal lattice
  {B}oltzmann solver using the {TheLMA} framework}.
\newblock In: \bibinfo{booktitle}{Proceedings of the 23st International
  Conference on Parallel Computational Fluid Dynamics}. Parallel CFD 2011;
  \bibinfo{year}{2011}{\natexlab{b}},.
%Type = Book
\bibitem[{Wolf-Gladrow(2000)}]{lba:wolf-gladrow:2000}
\bibinfo{author}{Wolf-Gladrow\xfnm[ D.A.]}.
\newblock \bibinfo{title}{Lattice-Gas Cellular Automata and Lattice {B}oltzmann
  Models}; vol. \bibinfo{volume}{1725} of \emph{\bibinfo{series}{Lecture Notes
  in Mathematics}}.
\newblock \bibinfo{address}{Berlin}: \bibinfo{publisher}{Springer};
  \bibinfo{year}{2000}.
%Type = Book
\bibitem[{Succi(2001)}]{lba:succi:2001b}
\bibinfo{author}{Succi\xfnm[ S.]}.
\newblock \bibinfo{title}{The Lattice {B}oltzmann Equation -- For Fluid
  Dynamics and Beyond}.
\newblock \bibinfo{publisher}{Clarendon Press}; \bibinfo{year}{2001}.
%Type = Article
\bibitem[{Chen and Doolen(1998)}]{lba:chen:1998}
\bibinfo{author}{Chen\xfnm[ S.]}, \bibinfo{author}{Doolen\xfnm[ G.D.]}.
\newblock \bibinfo{title}{Lattice {B}oltzmann method for fluid flows}.
\newblock \bibinfo{journal}{Annu Rev Fluid Mech}
  \bibinfo{year}{1998};\bibinfo{volume}{30}:\bibinfo{pages}{329--364}.
%Type = Article
\bibitem[{Qian et~al.(1992)Qian, d{'}Humi{\`e}res and
  Lallemand}]{lba:qian:1992}
\bibinfo{author}{Qian\xfnm[ Y.H.]}, \bibinfo{author}{d{'}Humi{\`e}res\xfnm[
  D.]}, \bibinfo{author}{Lallemand\xfnm[ P.]}.
\newblock \bibinfo{title}{Lattice {BGK} models for {N}avier-{S}tokes equation}.
\newblock \bibinfo{journal}{Europhys Lett}
  \bibinfo{year}{1992};\bibinfo{volume}{17}(\bibinfo{number}{6}):\bibinfo{page%
s}{479--484}.
%Type = Article
\bibitem[{Ziegler(1993)}]{lba:ziegler:1993}
\bibinfo{author}{Ziegler\xfnm[ D.P.]}.
\newblock \bibinfo{title}{Boundary conditions for lattice {B}oltzmann
  simulations}.
\newblock \bibinfo{journal}{J Stat Phys}
  \bibinfo{year}{1993};\bibinfo{volume}{71}(\bibinfo{number}{5/6}):\bibinfo{pa%
ges}{1171--1177}.
%Type = Misc
\bibitem[{zce(2007)}]{zcell:ibm:arch:2007}
\bibinfo{title}{{T}he {C}ell {B}roadband {E}ngine {A}rchitecture, {V}ersion
  1.02}.
\newblock \bibinfo{howpublished}{\newline
  \url{http://www.ibm.com/developerworks/power/cell/documents.html}};
  \bibinfo{year}{2007}.
%Type = Misc
\bibitem[{hpc(2011{\natexlab{a}})}]{hpc:AMD6970:2011}
\bibinfo{title}{{AMD} 6970 {GPU}}.
\newblock
  \bibinfo{howpublished}{\url{http://www.amd.com/us/products/desktop/graphics/%
amd-radeon-hd-6000/hd-6970/}}; \bibinfo{year}{2011}{\natexlab{a}}.
%Type = Article
\bibitem[{Habich et~al.(2011)Habich, Zeiser, Hager and
  Wellein}]{gpu:habich:ades}
\bibinfo{author}{Habich\xfnm[ J.]}, \bibinfo{author}{Zeiser\xfnm[ T.]},
  \bibinfo{author}{Hager\xfnm[ G.]}, \bibinfo{author}{Wellein\xfnm[ G.]}.
\newblock \bibinfo{title}{{P}erformance analysis and optimization strategies
  for a {D3Q19} lattice {B}oltzmann kernel on n{VIDIA} {GPU}s using {CUDA}}.
\newblock \bibinfo{journal}{Advances in Engineering Software}
  \bibinfo{year}{2011};\bibinfo{volume}{42}(\bibinfo{number}{5}):\bibinfo{page%
s}{266 -- 272}.
\newblock
  \urlprefix\url{http://www.sciencedirect.com/science/article/pii/S09659978100%
01274}.
%Type = Misc
\bibitem[{hpc(2011{\natexlab{b}})}]{hpc:OpenMP:2011}
\bibinfo{title}{{O}pen {MP}}.
\newblock \bibinfo{howpublished}{\url{http://www.openmp.org/}};
  \bibinfo{year}{2011}{\natexlab{b}}.
%Type = Misc
\bibitem[{gpu(2011)}]{gpu:nvidiaCudaProgrammingGuide4.0:2011}
\bibinfo{title}{{n}{VIDIA} {C}uda {P}rogramming {G}uide 4.0}.
\newblock
  \bibinfo{howpublished}{\url{http://developer.download.nvidia.com/compute/Dev%
Zone/docs/html/C/doc/CUDA_C_Programming_Guide.pdf}}; \bibinfo{year}{2011}.
%Type = Misc
\bibitem[{hpc(2011{\natexlab{c}})}]{hpc:KHRONOS:OpenCL:2011}
\bibinfo{title}{{KHRONOS GROUP (TM)} {OpenCL} {Specification}}.
\newblock \bibinfo{howpublished}{\url{http://www.khronos.org/opencl/}};
  \bibinfo{year}{2011}{\natexlab{c}}.
%Type = Inproceedings
\bibitem[{Volkov and Demmel(2008)}]{Volkov:2008:BGT:1413370.1413402}
\bibinfo{author}{Volkov\xfnm[ V.]}, \bibinfo{author}{Demmel\xfnm[ J.W.]}.
\newblock \bibinfo{title}{Benchmarking {GPU}s to tune dense linear algebra}.
\newblock In: \bibinfo{booktitle}{Proceedings of the 2008 ACM/IEEE conference
  on Supercomputing}. SC '08; \bibinfo{address}{Piscataway, NJ, USA}:
  \bibinfo{publisher}{IEEE Press}.
\newblock ISBN \bibinfo{isbn}{978-1-4244-2835-9}; \bibinfo{year}{2008}, p.
  \bibinfo{pages}{31:1--31:11}.
\newblock
  \urlprefix\url{http://portal.acm.org/citation.cfm?id=1413370.1413402}.
%Type = Misc
\bibitem[{hpc(2011{\natexlab{d}})}]{hpc:StreamBench:2011}
\bibinfo{title}{{T}he {S}tream benchmark}.
\newblock \bibinfo{howpublished}{\url{http://www.streambench.org/}};
  \bibinfo{year}{2011}{\natexlab{d}}.
%Type = Misc
\bibitem[{hpc(2011{\natexlab{e}})}]{hpc:Wiki:ECC}
\bibinfo{title}{{Wikipedia} {ECC-Memory}}.
\newblock
  \bibinfo{howpublished}{\url{http://en.wikipedia.org/wiki/ECC_memory}};
  \bibinfo{year}{2011}{\natexlab{e}}.
%Type = Article
\bibitem[{Wellein et~al.(2006)Wellein, Zeiser, Hager and
  Donath}]{2006:Wellein:CompFluids}
\bibinfo{author}{Wellein\xfnm[ G.]}, \bibinfo{author}{Zeiser\xfnm[ T.]},
  \bibinfo{author}{Hager\xfnm[ G.]}, \bibinfo{author}{Donath\xfnm[ S.]}.
\newblock \bibinfo{title}{{On the single processor performance of simple
  lattice Boltzmann kernels}}.
\newblock \bibinfo{journal}{Computers \& Fluids}
  \bibinfo{year}{2006};\bibinfo{volume}{35}(\bibinfo{number}{8-9}):\bibinfo{pa%
ges}{910--919}.
%Type = Article
\bibitem[{Feichtinger et~al.(2011{\natexlab{a}})Feichtinger, Donath,
  K{\"o}stler, G{\"o}tz and R{\"u}de}]{hpc:Feichtinger:2011}
\bibinfo{author}{Feichtinger\xfnm[ C.]}, \bibinfo{author}{Donath\xfnm[ S.]},
  \bibinfo{author}{K{\"o}stler\xfnm[ H.]}, \bibinfo{author}{G{\"o}tz\xfnm[
  J.]}, \bibinfo{author}{R{\"u}de\xfnm[ U.]}.
\newblock \bibinfo{title}{{W}alberla: {HPC} software design for computational
  engineering simulations}.
\newblock \bibinfo{journal}{Journal of Computational Science}
  \bibinfo{year}{2011}{\natexlab{a}};\bibinfo{volume}{In
  Press,}(\bibinfo{number}{2008}).
\newblock
  \urlprefix\url{http://linkinghub.elsevier.com/retrieve/pii/S1877750311000111%
}.
%Type = Article
\bibitem[{Feichtinger et~al.(2011{\natexlab{b}})Feichtinger, Habich,
  K{\"o}stler, Hager, R{\"u}de and Wellein}]{Feichtinger:2010:ParCo}
\bibinfo{author}{Feichtinger\xfnm[ C.]}, \bibinfo{author}{Habich\xfnm[ J.]},
  \bibinfo{author}{K{\"o}stler\xfnm[ H.]}, \bibinfo{author}{Hager\xfnm[ G.]},
  \bibinfo{author}{R{\"u}de\xfnm[ U.]}, \bibinfo{author}{Wellein\xfnm[ G.]}.
\newblock \bibinfo{title}{A flexible patch-based lattice {B}oltzmann
  parallelization approach for heterogeneous {GPU--CPU} clusters}.
\newblock \bibinfo{journal}{Parallel Computing}
  \bibinfo{year}{2011}{\natexlab{b}};\bibinfo{volume}{37}(\bibinfo{number}{9})%
:\bibinfo{pages}{536 -- 549}.
\newblock
  \urlprefix\url{http://www.sciencedirect.com/science/article/pii/S01678191110%
00342}.
%Type = Article
\bibitem[{Donath et~al.(2011)Donath, Mecke, Rabha, Buwa and
  R{\"u}de}]{2010:Donath:CompFlu}
\bibinfo{author}{Donath\xfnm[ S.]}, \bibinfo{author}{Mecke\xfnm[ K.]},
  \bibinfo{author}{Rabha\xfnm[ S.]}, \bibinfo{author}{Buwa\xfnm[ V.]},
  \bibinfo{author}{R{\"u}de\xfnm[ U.]}.
\newblock \bibinfo{title}{{Verification of Surface Tension in the Parallel Free
  Surface Lattice Boltzmann Method in WaLBerla}}.
\newblock \bibinfo{journal}{Computers \& Fluids}
  \bibinfo{year}{2011};\bibinfo{volume}{2}(\bibinfo{number}{2}):\bibinfo{pages%
}{105 -- 112}.
%Type = Article
\bibitem[{G{\"o}tz et~al.(2010)G{\"o}tz, Iglberger, St{\"u}rmer and
  R{\"u}de}]{2010:Goetz:SC}
\bibinfo{author}{G{\"o}tz\xfnm[ J.]}, \bibinfo{author}{Iglberger\xfnm[ K.]},
  \bibinfo{author}{St{\"u}rmer\xfnm[ M.]}, \bibinfo{author}{R{\"u}de\xfnm[
  U.]}.
\newblock \bibinfo{title}{Direct numerical simulation of particulate flows on
  294912 processor cores}.
\newblock \bibinfo{journal}{IEEE computer society (Veranst): 2010 ACM/IEEE
  International Conference for High Performance Computing, Networking, Storage
  and Analysis (Supercomputing 2010, New Orleans, 1311 -- 19112010)}
  \bibinfo{year}{2010};:\bibinfo{pages}{1 -- 11}.
%Type = Inproceedings
\bibitem[{Wellein et~al.(2011)Wellein, Habich, Hager and
  Zeiser}]{Wellein:ParCFD11}
\bibinfo{author}{Wellein\xfnm[ G.]}, \bibinfo{author}{Habich\xfnm[ J.]},
  \bibinfo{author}{Hager\xfnm[ G.]}, \bibinfo{author}{Zeiser\xfnm[ T.]}.
\newblock \bibinfo{title}{{Node-level} performance of the lattice {B}oltzmann
  method on recent multicore {CPUs}}.
\newblock In: \bibinfo{booktitle}{Proceedings of the 23st International
  Conference on Parallel Computational Fluid Dynamics}. Parallel CFD 2011;
  \bibinfo{year}{2011},.
%Type = Article
\bibitem[{Treibig et~al.(2010)Treibig, Hager and
  Wellein}]{10.1109/ICPPW.2010.38}
\bibinfo{author}{Treibig\xfnm[ J.]}, \bibinfo{author}{Hager\xfnm[ G.]},
  \bibinfo{author}{Wellein\xfnm[ G.]}.
\newblock \bibinfo{title}{Likwid: A lightweight performance-oriented tool suite
  for x86 multicore environments}.
\newblock \bibinfo{journal}{Parallel Processing Workshops, International
  Conference on Parallel Processing Workshops}
  \bibinfo{year}{2010};\bibinfo{volume}{0}:\bibinfo{pages}{207--216}.
%Type = Misc
\bibitem[{hpc(2011{\natexlab{f}})}]{hpc:likwid}
\bibinfo{title}{{LIKWID}}.
\newblock \bibinfo{howpublished}{\url{http://code.google.com/p/likwid}};
  \bibinfo{year}{2011}{\natexlab{f}}.

\end{thebibliography}

%% not want to use model3-num-names.bst.





% \end{thebibliography}

\end{document}